\documentclass[9 pt,twocolumn]{revtex4}

\usepackage{color}
\usepackage{graphicx}
\usepackage{dcolumn}
\usepackage{bm}
\usepackage{hyperref}
\usepackage{amssymb}
\usepackage{hyperref}
\begin{document}
\newcommand{\be}{\begin{equation}}
\newcommand{\ee}{\end{equation}}
\newcommand{\ba}{\begin{eqnarray}}
\newcommand{\ea}{\end{eqnarray}}
\newcommand{\Gam}{\Gamma[\varphi]}
\newcommand{\Gamm}{\Gamma[\varphi,\Theta]}
\thispagestyle{empty}

\title{Entanglement in $S$ states of two-electron quantum dots with   Coulomb
impurities at the center.
  }
\author{ Przemys\l aw Ko\'scik,
Institute of Physics,  Jan Kochanowski University\\
ul. \'Swi\c{e}tokrzyska 15, 25-406 Kielce, Poland}

 \begin{abstract}
 We study a  system of two Coulombically interacting
electrons in an external harmonic
 potential in the presence of an on-centre Coulomb impurity.
 Detailed
results for the dependencies of the reduced von Neumann entropy on
the control parameters of the system  are provided for both the
ground state and  the triplet  $S$ states with the lowest energy.
Among other features, it is found that in the weak confinement
regime the entanglement is   strongly affected by the presence of an
acceptor impurity.

 \end{abstract}

\maketitle

\section{Introduction}

In the last few years there has been an explosion of interest in
the entanglement properties of systems  of interacting
particles, in view of  their possible use in quantum information
technology \cite{inf,inf1}. In particular, great efforts have been
made to explore quantum entanglement in systems of two
interacting particles, including model systems such as the
Moshinsky atom \cite{1,2,3,3.1},  the Crandall atom \cite{4}, or
quantum dot systems with harmonically shaped traps \cite{5,6,7,8,9}.
 In the most recent years, research
activity has expanded towards investigating  the entanglement  in
real two-electron
 systems, i.e., the helium atoms and helium ions.  For   instance, Manzano \textsl{et al.} \cite{4}, Dehesa
\textsl{et al.}\cite{10} and, Benenti \textsl{et al}.\cite{11} have
addressed
  their investigations  to  the  relation  between
entanglement and energy for the helium atom.  Lin \textsl{et al.}
\cite{12} calculated the linear entropy for the helium ion
$|1sns;^{1}S\rangle$ states for $n$ up to $10$, for  a wide range
of values of the nuclear charge (from $Z=2$ to $15$).
They found that  in contrast to the
ground state, the spatial entanglement of excited states increases
with increasing nuclear charge. The entanglement  in the spherical
helium model, where the Coulombic interaction between the electrons is
replaced by its spherical average, has also been discussed in the
literature \cite{en1,en2}. An overview of  recent developments, both theoretical
and experimental, in entanglement studies of  quantum
composite systems, including atoms and molecules, can be found in
\cite{tishy}.

 Few attempts have been
 made recently towards investigating the effect of a Coulomb impurity on the
properties of quantum dots \cite{im,im2,im5,im6}. The systems
of  confined helium atoms and helium ions that are closely related
to quantum dots with a Coulomb impurity have received much
attention in recent years \cite{13,15,16,17,18,19}. However, as far
as we know,  the entanglement properties of such systems have not
been extensively studied in the literature. A simple candidate for
studying the Coulomb impurity effect is the system
 of two  interacting electrons in the dot
 modeled by a harmonic potential
\begin{eqnarray} H=\sum_{i=1}^{2}[-{\hbar^2\over 2m}
\bigtriangledown_{i}^{2}+{m\omega^2\over 2}
r_{i}^2+{e^2\over\varepsilon}{\eta\over r_{i}}]+{e^2\over
\varepsilon}{1\over r_{12}},\label{hamff}\end{eqnarray} where $\eta$
is the effective charge of the  impurity. In particular, for
$\eta=-2,-1$, the system (\ref{hamff}) describes  the confined helium
$He$ atoms and helium ions $H^{-}$ centred in a harmonic trap,
respectively. The effects of the harmonic confinement  on the ground
state energy of a helium atom was estimated for a series of values
of $\omega$ in \cite{15}.

The purpose of the present Letter is to gain  insight into how both
the effective charge and the confinement size influence the
entanglement. To fully reveal the effect of the Coulomb impurity on
the entanglement, we provide results for both the acceptor
($\eta<0$) and donor ($\eta>0$) impurities and compare them with the
ones for impurity-free dots ($\eta=0$). In particular, we study both
$He$ and $H^{-}$ to study the confinement effect on their
entanglement properties.

The scaling $\textbf{r}\rightarrow\sqrt{{\hbar\over
m\omega}}\textbf{r}$, $E\rightarrow \hbar \omega E$ turns the
Schr\"{o}dinger equation into \be
H\psi(\textbf{r}_{1},\textbf{r}_{2})=
E\psi(\textbf{r}_{1},\textbf{r}_{2}),\label{shrodi}\ee with \be
H=\sum_{i=1}^{2}[-{1\over 2} \bigtriangledown_{i}^{2}+{1\over 2}
r_{i}^2+{ \eta \gamma\over r_{i}}]+{\gamma\over r_{12}},
\label{ham}\ee  where $\gamma={e^2\over\varepsilon}\sqrt{{m\over
\omega\hbar^{3}}}$. The limits as $\gamma\rightarrow0$ and
$\gamma\rightarrow\infty$ correspond  to situations in which the
frequency $\omega$  of the trap tends to $\infty$ and $0$,
respectively.  In the latter limit we recover a free space impurity
 system (atomic-like system), which has at least one bound state for $\eta\leq
\eta_{cr}\approx -0.911$ \cite{critical}; for $\eta_{cr}<\eta<0$, it
consists of one electron bound to the impurity charge and one
free outer electron.

 Since the Hamiltonian
(\ref{ham}) does not contain any terms that couple the spatial and
the spin coordinates,  its stationary states possess the form
 \be \Psi_{S\atop
T}(\xi_{1},\xi_{2})=\psi^{\pm}(\textbf{r}_{1},\textbf{r}_{2})\chi_{S\atop
T}^{s_{z}}, \label{fact} \ee where $\chi_{S\atop T}^{s_{z}}$ are the
spin functions and the spatial wavefunctions  are symmetric $(+)$ or
antisymmetric $(-)$ under permutation of the electrons.

In this Letter, we restrict our investigation to the $S$ states, the
spatial  wavefunctions  which
 depend explicitly only on the radial coordinates $r_{1}, r_{2}$ and $\cos\theta$, where $\theta$ is the inter-electronic
angle coordinate,
$\psi^{\pm}(\textbf{r}_{1},\textbf{r}_{2})\equiv\psi^{\pm}(r_{1},r_{2},\cos\theta)$
and the differential volume element is $d\tau = 8\pi^2 \sin\theta
r_{1}^2r_{2}^2dr_{1} dr_{2}d\theta$ \cite{hyleras}.

This Letter is organized as follows. In Section \ref{12}, we discuss
an effective procedure  to analyze the entanglement of the singlet
and triplet $S$-symmetry states.
 Section \ref{res} is devoted
to the results  and some concluding remarks are left for Section
\ref{summ}.

\section{Entanglement of $S$-states }\label{12}

The real spatial wavefunction $\psi^{\pm}(r_{1},r_{2},\cos\theta)$
can be expanded in a Fourier-Legendre series of Legendre polynomials
of the cosine of $\theta$\cite{kaplan}:
\begin{eqnarray}
\psi^{\pm}(\textbf{r}_{1},\textbf{r}_{2})=\sum_{l=0}^{\infty}{f_{l}^{\pm}(r_{1},r_{2})\over
r_{1}r_{2}}P_{l}(\cos\theta),\label{hjk}\end{eqnarray} where
 \begin{eqnarray}
f_{l}^{\pm}(r_{1},r_{2})=\nonumber\\=r_{1} r_{2}{2l+1\over 2}
\int_{0}^{\pi}\psi^{\pm
}(r_{1},r_{2},\cos\theta)P_{l}(\cos\theta)\sin\theta
d\theta.\label{plp}
\end{eqnarray}
Being real and symmetric, the function  $f_{l}^{+}(r_{1},r_{2})$ has
the Schmidt decomposition \cite{vn1}
\begin{eqnarray}
 f_{l}^{+}(r_{1},r_{2})=\sum_{n=0}^{\infty} k_{nl}^{+} v_{nl}^{+} (r_{1})
v_{nl}^{+}(r_{2}),\label{sym}\end{eqnarray} where the orbitals
$v_{nl}^{+}$ satisfy $\langle
v_{nl}^{+}|v_{n^{'}l}^{+}\rangle=\delta_{nn^{'}}$. On the other
hand, the
  function $f_{l}^{-}(r_{1},r_{2})$  is real and antisymmetric
 and so its Slater decomposition  \cite{vn1} is
\begin{eqnarray}
 f_{l}^{-}(r_{1},r_{2})=\nonumber\\=\sum_{n=0}^{\infty}k_{nl}^{-} ( v_{2n+1l}^{-} (r_{1})v_{2nl}^{-}
 (r_{2})
-v_{2n+1l}^{-} (r_{2})v_{2nl}^{-}
(r_{1})),\label{anty}\end{eqnarray}  where $\langle
v_{nl}^{-}|v_{n^{'}l}^{-} \rangle=\delta_{nn^{'}}$. With the help of
the  expansions
 (\ref{sym}) and (\ref{anty}), the addition theorem $
P_{l}(\cos\theta)={4 \pi\over 2l+1}
\sum_{m=-l}^{l}Y_{lm}^{*}(\theta_{1},\varphi_{1})Y_{lm}(\theta_{2},\varphi_{2})
$, and the identity $[Y_{l,m}(\theta,\varphi)]^{*}=(-1)^m
Y_{l,-m}(\theta,\varphi)$, the   Slater decompositions of the total
singlet and triplet $S$-symmetry wavefunctions (\ref{fact}) can be
easily inferred. The number of non-zero expansion coefficients in
the Slater decomposition is called the Slater rank (SR). An
essential point is that a pure fermion state is non-entangled if,
and only if, it can be expressed by a single Slater determinant
\cite{ghirardi}, i.e., its SR is equal to one.

Quantum entanglement is characterized by the
 spectrum of the single-particle reduced density matrix \cite{rdm},
$\rho_{red}=Tr_{2}[|\Psi\rangle\langle\Psi|]$, and many ways of
measuring its amount have been developed \cite{
vn1,ghirardi,par,vn31,lin0,vn2}. In this Letter, to quantify the
entanglement  we shall use  the reduced  von Neumann (vN) entropy
 \be S_{vN}=S[\rho_{red}]-1,\label{polo}\ee  where
$S[\rho_{red}]=Tr[\rho_{red}\log_{2}\rho_{red}]$ is the ordinary vN
entropy \cite{vn1}. The above measure  vanishes for a two fermion
pure state when its total wavefunction can be expressed as one
single determinant \cite{vn2}. Since the total wavefunction
factorizes into spatial and spin components, the same holds for the
reduced density matrix
$\rho_{red}=\rho^{spatial}_{red}\rho^{spin}_{red}$ and, in
consequence, $S[\rho_{red}]$ separates into
$S[\rho_{red}]=S[\rho^{spatial}_{red}]+S[\rho^{spin}_{red}]$, where
the spin contribution  depends only on $|s_{z}|$ \cite{vn3}, that
is, $S[\rho^{spin}_{red}]=-|s_{z}|+1$. Therefore, the reduced vN
entropy can be written as \be S_{S\atop
T}^{s_{z}=\sigma}=-|\sigma|-Tr[\rho^{\pm}_{red}\log_{2}
\rho^{\pm}_{red}],\label{ddd}  \ee where $\rho^{\pm}_{red}$ are the
spatial reduced density matrices of the
 singlet
 and triplet  states,
\be \rho^{\pm}_{red}(\textbf{r},\textbf{r}^{'})=\int [\psi^{
\pm}(\textbf{r},\textbf{r}_{1})]^{*}\psi^{
\pm}(\textbf{r}^{'},\textbf{r}_{1})d\textbf{r}_{1}.\ee

 The
eigenvalues (occupancies) $\lambda_{nl}^{+}$ and $\lambda_{nl}^{-}$
of the spatial reduced density matrices of the
 singlet
 and triplet  states are related
to the coefficients $k_{nl}^{\pm}$ by
  $\lambda_{nl}^{\pm}= ({4\pi k_{nl}^{\pm}\over 2l+1})^2$.
The former   are $2l+1$-fold degenerate, whereas the latter are
$2(2l+1)$-fold degenerate, so that  the normalization conditions
give
$\sum_{nl}(2l+1)\lambda_{nl}^{+}=2\sum_{nl}(2l+1)\lambda_{nl}^{-}=1$.
As the vN entropies of the triplet states
  with $s_{z}=0$ and with $s_{z}=\pm1$
 differ from each other only by one, we will concentrate mainly on
$S_{T}^{s_{z}=\pm1}$, without loss of generality. In terms of the
occupancies, the vN entropies (\ref{ddd}) of the singlet and the
triplet with ${s_{z}=\pm {1}}$ states are given by
  $S_{S}=-\sum_{nl} (2l+1)
\lambda_{nl}^{+} \log_{2} \lambda_{nl}^{+}$ and
$S^{s_{z}=\pm{1}}_{T}=-1-2\sum_{nl} (2l+1)\lambda_{nl}^{-} \log_{2}
\lambda_{nl}^{-}$, respectively.

The radial orbitals $v_{nl}^{\pm}$ satisfy the following integral
equations \be \int_{0}^{\infty}
\rho^{\pm}_{l}(r,r^{'})v_{pl}^{\pm}(r^{'})dr^{'}=[k_{nl}^\pm]^2
v_{pl}^{\pm}(r), \label{nn}\ee with
$\rho^{\pm}_{l}(r,r^{'})=\int_{0}^{\infty} f^{\pm}_{l} (r,r_{1})
f^{\pm}_{l} (r^{'},r_{1}) dr_{1}$, where $p=n$ and  $p=2n, 2n+1$ for
$\rho^{+}_{l}$ and $\rho^{-}_{l}$, respectively. With the help of
Eq. (\ref{plp}), we find that $\rho^{\pm}_{l}$ can be expressed by
the following $3$-dimensional integrals.
\begin{eqnarray} \rho^{\pm}_{l}(r,r^{'})=r r^{'}({2l+1\over
2})^{2}\nonumber\\\times \int_{0}^{\infty}\int
_{0}^{\pi}\int_{0}^{\pi}r_{1}^2 \psi^{\pm
}(r,r_{1},\cos\theta)P_{l}(\cos\theta)\sin\theta
\nonumber\\\times\psi^{\pm
}(r^{'},r_{1},\cos\theta^{'})P_{l}(\cos\theta^{'})\sin\theta^{'}
dr_{1}d\theta d\theta^{'}.\label{rdm1}\end{eqnarray}
 In
particular,  the  orbitals $v_{nl}^{+}$ and their eigenvalues
$k_{nl}^+$ can be determined by \be \int_{0}^{\infty}
f_{l}^{+}(r,r^{'}) v_{nl}^{+} (r^{'}) dr^{'}=k_{nl}^{+}
v_{nl}^{+}(r).\label{ppol1}\ee One of the most efficient ways to
determine  $\lambda_{nl}^{\pm}$ is to solve  the above integral
equations through a discretization technique (see for example
\cite{6}). Since it is easier to obtain $f^{+}_{l}$ (\ref{plp}) than
$\rho^{+}_{l}$ (\ref{rdm1}), we will deal throughout this Letter
with
 (\ref{ppol1}) when  the $\lambda_{nl}^{+}$ are needed.

\section{Numerical results and discussion}\label{res}

 To compute the $S$-state energies and wavefunctions of  (\ref{ham}), we use a simple but effective variational wavefunction
  \be \psi_{Hyl}= e^{-\mu s } \sum_{nmp}
c_{nmp}s^{n}t^{m}u^{p},\label{funm}\ee   where $s, t$, and $u$ are
the Hylleraas coordinates;
 $s=r_{1}+r_{2}, t=r_{2}-r_{1}, u=r_{12}$, and $\mu$ is a non-linear
 variational parameter \cite{hyleras}.  In order to gain insight into the
effectiveness of the method, we  first determine the occupancies of
 the ground state of the free space impurity system with $\eta=-2$
 (helium) and
  assess their accuracy by comparing the linear entropy $L=1-\sum_{nl}(2l+1) [\lambda_{nl}^{+}]^2$  with the already available data in the
literature, $L= 0.015 914$ \cite{10}, 0.01606 \cite{11}, 0.015943
\cite{12}.
 We use an expansion given by a $372$-term wavefunction (\ref{funm}) that includes all terms consistent with the
condition $0\leq n+m+p\leq 14$, ($m$ even), reproducing at $\mu=3$
the ground-state helium  energy with to at least ten significant
digits: $-2.903724377$ \cite{hl}. Discretizing the variables $r$ and
$r^{'}$ with equal subintervals of length $\bigtriangleup r ={R/
n_{max}}$, we
 turn (\ref{ppol1}) into an algebraic eigenvalue problem\be
\sum_{i=0}^{n_{max}}[M_{ij}^{(l)}-\delta_{ij}k_{nl}^{+}]v_{nl}^{+}(\bigtriangleup
r i)=0, j,\label{pp}\ee where $M_{ij}=\bigtriangleup r
f^{+}_{l}(\bigtriangleup r i,\bigtriangleup r j) $. Diagonalization
of the matrix $[M_{ij}^{(l)}]$ yields thus a set of approximations to
the coefficients $k_{nl}^{+}$. With the occupancies  determined in
such a way that $\lambda_{nl}^{+}= ({4\pi k_{nl}^{+}\over 2l+1})^2$,
we can then obtain
 an  approximation to the true value of $L$, $L_{ap}=1-\sum_{n,l=0}^{n_{max},l_{max}}(2l+1) [\lambda_{nl}^{+}]^2$. For illustrative purposes, we present in  Table
\ref{tab:1kku}  the  values obtained for $L_{ap}$ with
different    cut-offs $l_{max}$ and $n_{max}$, where  
  we obtain the
value
  $ 0.0159172$ for the linear entropy,  in excellent agreement with
the results of Refs. \cite{10,11,12} wherein they were obtained in
different ways.

\begin{table}[h]
\begin{center}
\begin{tabular}{|l|l|l|l|}
\hline
$$ & $n_{max}=100$ &$n_{max}=200$&$n_{max}=300$ \\
\hline
$ l_{max}=0$ & $0.0160148$ & $0.0159268$&$0.0159221$ \\

$ l_{max}=1$ & $0.0160100$ & $0.0159220$&$0.0159172$ \\

\hline
\end{tabular}
\caption{\label{tab:1kku} $L_{ap}$ computed with different $l_{max}$
and $n_{max}$, for $R=7.5$ where the functions $f^{+}_{l}$ are
mainly confined.}
\end{center}
\vspace{-0.6cm}
\end{table}

\begin{figure}[h]
\begin{center}
\end{center}
\includegraphics[width=0.46\textwidth]{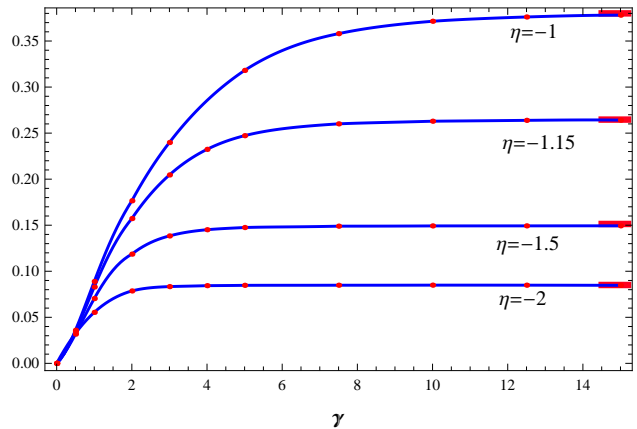}

\caption{ The von Neumann entropy  of the ground-state as a function
of $\gamma$ for some different  exemplary values of $\eta$
($\eta\leq \eta_{cr}$). The horizontal lines mark the results for
the free space impurity systems.} \label{fig:odog1}
\end{figure}

Now we come to the main goal of this Letter, which is to explore the
entanglement in the ground state and the triplet $S$ states of
lowest energy of (\ref{ham}). First we treat the systems in the
presence of  acceptor impurities $(\eta<0)$.
  For all negative charge $\eta$ values  considered in
this Letter, one has that $l$ up to $l_{max}=4$ is typically
sufficient to get a good estimate of the  vN entropy, over the
entire range of values of $\gamma$.
 Fig. \ref{fig:odog1} depicts the dependence of the ground-state vN entropy $S_{S}$ on  $\gamma$
  for  some
 exemplary values of $\eta$ smaller than $\eta_{cr}$. In particular, the results of this figure reveal  the
 effects
  of the confinement on the entanglement in the ground states of the
  helium $He$ atom
 and the helium ion $H^{-}$
  (the cases $\eta=-2$ and $\eta=-1$, respectively).\begin{figure}[h]
\includegraphics[width=0.46\textwidth]{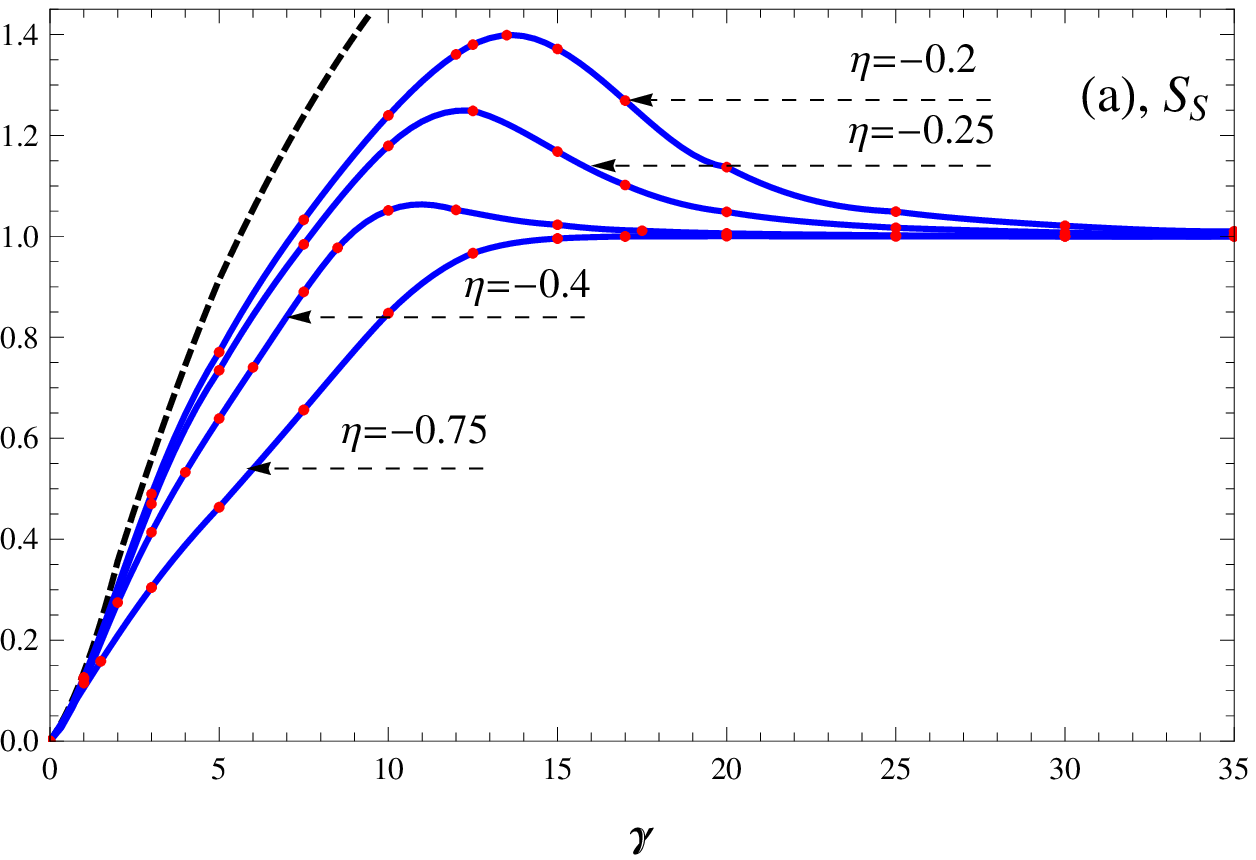}
\includegraphics[width=0.46\textwidth]{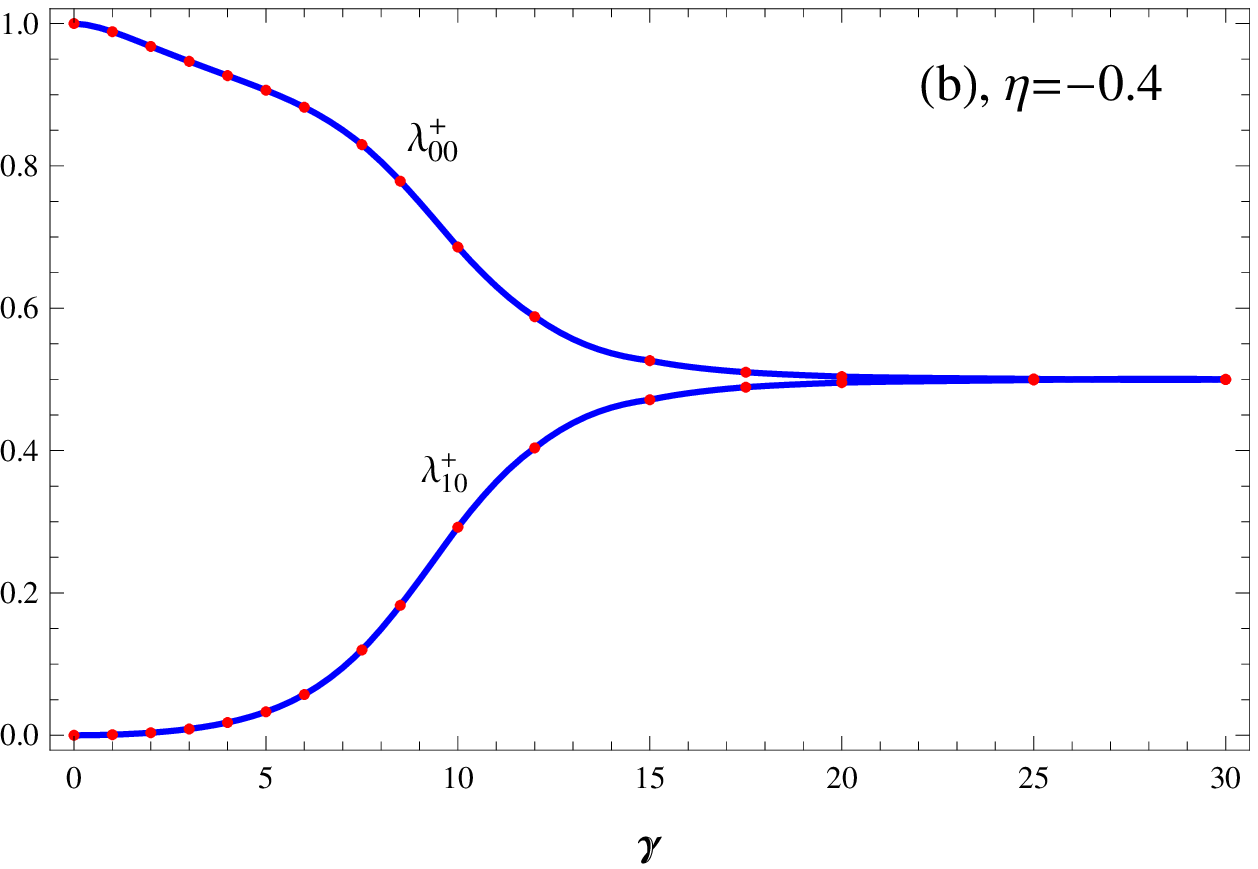}
\begin{center}
\end{center}
\caption{(a) The von Neumann entropy  of the  ground state as a
function of $\gamma$ for some different negative values of $\eta$
($\eta>\eta_{cr}$). The black-dashed line corresponds to the case
$\eta=0$.
  (b) The behaviour of the two largest occupancies determined as functions of $\gamma$ for
$\eta=-0.4$. }  \label{fig:odg}
\end{figure} As far as
  we know,  in none of the papers concerning
confined helium ions have the results for their entanglement
properties appeared.  In the limit of
$\gamma\rightarrow\infty$, the vN entropy $S_{S}$ approaches
strictly a value of the corresponding state of the free space
impurity system, being smaller at smaller $\eta$. A  transition to
the free space impurity system regime is manifested by the onset of
the plateau in the behaviour of the vN entropy. The results of Fig.
\ref{fig:odog1} indicate that the critical value of $\gamma$ at which
this occurs decreases with decreasing $\eta$. This can be
qualitatively understood by referring to the localization of the
electrons, namely, for smaller $\eta$, they are  more localized
around the center and, in consequence, a stronger confinement  is
needed (smaller $\gamma$) to
 change their quantum state properties. The vN entropy deviates more and more    from
 the free space impurity value as the confinement becomes stronger and
 stronger. The  deviation is the largest
 in the limit of an infinitely strong confinement ($\gamma\rightarrow 0$), when the system
 behaves like non-interacting  electrons in a harmonic potential
 well, which gives  $S_{S}=0$ (SR=1).

  We also calculated the ionization threshold value $\gamma_ {c}$ for the confined helium atom, which is    defined as that for which
$E(He)=E({He}^{+})$, where $E(He)$ and $E({He}^{+})$ are the ground
state energies of $He$ and $He^{+}$ confined in the same
potential well\cite{17}. This means that for $\gamma =\gamma_ {c}$,
the confined helium atom gets ionized, i.e., it consists of one
electron bound to the nucleus and one unbound electron, but
still confined within the harmonic potential well.  Our highly
numerical result is $\gamma_{c}\approx0.832658$.  There may be a
general interest in noting that  there is no characteristic change
in the  behaviour of the vN entropy
 near
$\gamma_{c}$.

 When $\eta$ becomes larger than $\eta_{cr}$, the entanglement exhibits a qualitatively different behaviour from that of Fig. \ref{fig:odog1}; see
Fig. \ref{fig:odg} (a), namely the  vN entropy has a visibly
non-monotonic behaviour and in the limit of
$\gamma\rightarrow\infty$ saturates at a constant value that is
insensitive to $\eta$. The last point
 can be understood by referring to Ref. \cite{en1}
 wherein it was found that the  $|(1s)^2;^{1}S\rangle$ state  of  the free space impurity system, which corresponds to $\gamma\rightarrow\infty$, has for
$\eta_{cr}<\eta<0$
 only the two non-vanishing   occupancies $\lambda_{00}^{+}$ and $\lambda_{10}^{+}$, both equal to  ${1\over
 2}$, giving $S_{S}=1$ (SR=2). We recall here  that for
$\eta>\eta_{cr}$, the  $|(1s)^2;^{1}S\rangle$ state of the free
space impurity
 system is no longer a bound state.
 For the sake of
 illustration, Fig. \ref{fig:odg} displays in (b) the occupancies
$\lambda_{00}^{+}$ and $\lambda_{10}^{+}$  determined as functions
of $\gamma$ for the example of $\eta=-0.4$. It is seen how they
converge to an asymptotic doublet of the value ${1\over 2}$ when
$\gamma$ increases. Since both occupancies correspond to the purely
radial natural orbitals, the bulk of
 the entanglement is manifested  only in the radial variables as
 $\gamma\rightarrow\infty$. To gain a deeper insight into
 the effect of the Coulomb impurity, the variation of $S_{S}$ for the impurity-free dots ($\eta=0$)  is  also shown  in Fig. \ref{fig:odg} (a) (the black
dashed  line). The vN entropy $S_{S}$ with $\eta=0$ grows
monotonically with an increase in $\gamma$ and goes to infinity as
$\gamma\rightarrow\infty$, reflecting the fact that as the dot size
increases, more and more occupancies with different $l$ become
substantial \cite{cios}.
 \begin{figure}[h]
\begin{center}
\includegraphics[width=0.46\textwidth]{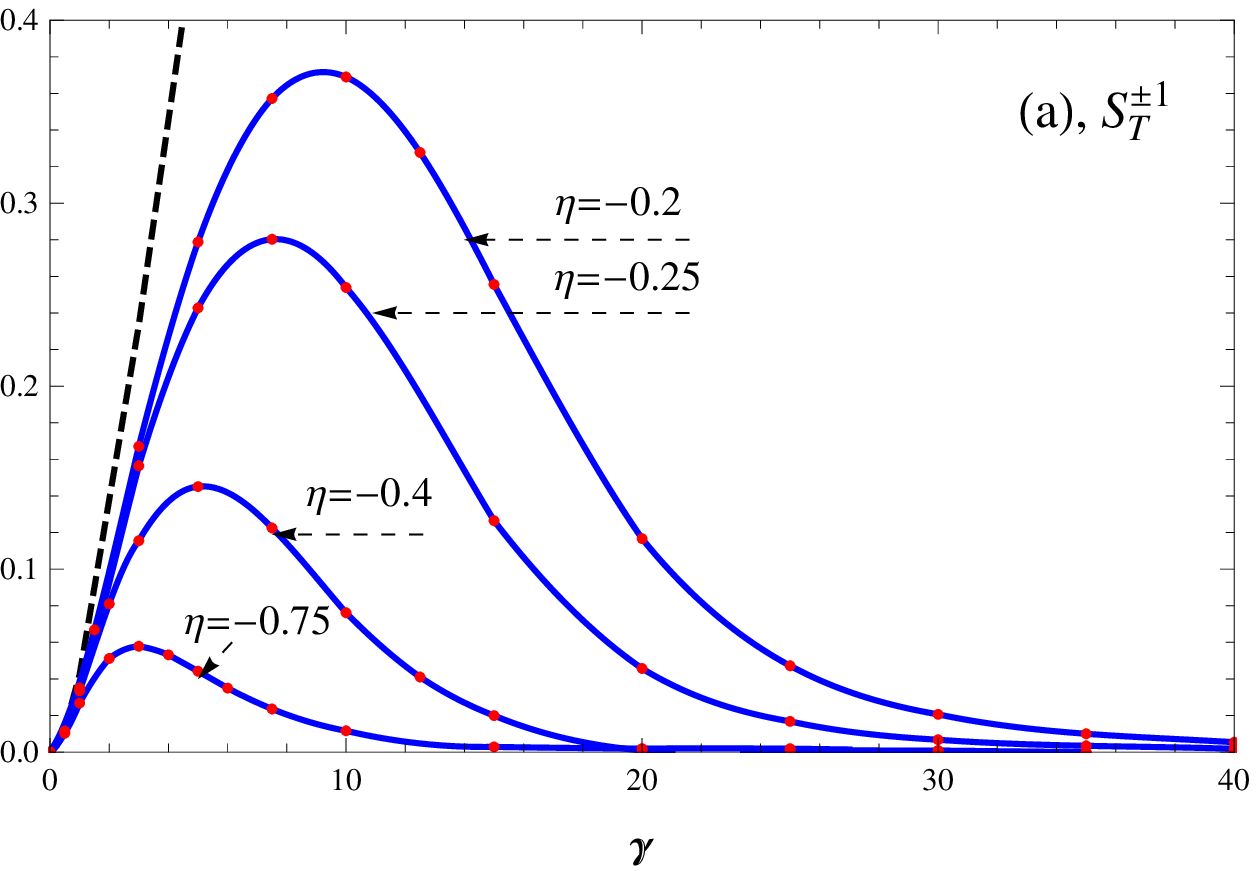}
\includegraphics[width=0.46\textwidth]{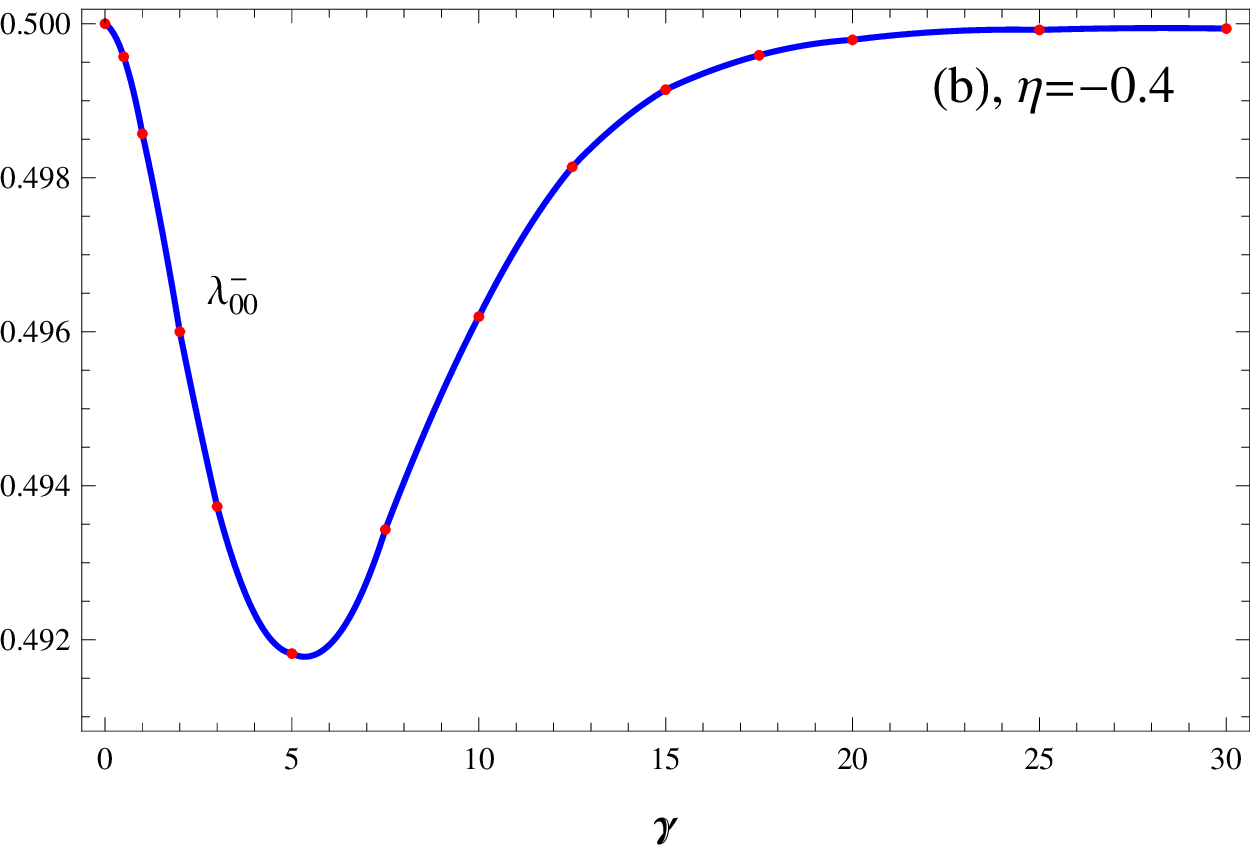}
\end{center}
\caption{(a) The von Neumann entropy $S_{T}^{s_{z}=\pm1}$ as a function
of $\gamma$ calculated for the same values of $\eta$ as in (a) of Fig.
 \ref{fig:odg}. The black-dashed line corresponds to the case $\eta=0$.  (b) Behaviour of the largest occupancy as a function of $\gamma$  for $\eta=-0.4$. }  \label{fig:ooooodg}
\end{figure}
\begin{figure}[h]
\begin{center}
\includegraphics[width=0.46\textwidth]{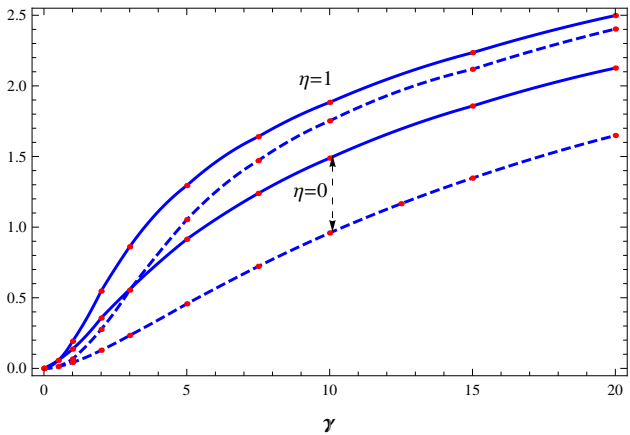}
\end{center}
\caption{Behaviour of the vN entropies $S_{S}$ (continuous
lines) and $S_{T}^{s_{z}=\pm1}$ (dashed lines),  for  $\eta=0, 1$ as
functions of  $\gamma$.} \label{fig:eeeeeodg1}
\end{figure}
 Interestingly enough, one sees that the vN entropy $S_{S}$ starts to exhibit   a clear
local maximum after exceeding a value of about $\eta\approx-0.4$.
The closer $\eta$
 is to zero, the more pronounced is the maximum and
the larger is the value of $\gamma$ at which it occurs. At the same
time, one observes that the range of values of $\gamma$ around
$\gamma=0$ in which $S_{S}$  exhibits the behaviour of the
impurity-free dots also  becomes wider.

  Fig.
\ref{fig:ooooodg} (a) reveals the effects of changing both $\eta$
and $\gamma$ on the entanglement in the lowest energy triplet state
with $s_{z}=\pm 1$, where for the sake of comparison the same values
of $\eta$ as those for the  ground state
 in Fig. \ref{fig:odg} (a) are taken into account. Once again, as for the singlet ground state, the black
dashed line represents  the case with no impurity. Except for this
case, for all remaining  cases   an analogous situation occurs,
namely the vN entropy increases, attains a maximum value, and then
 diminishes until it vanishes as $\gamma\rightarrow\infty$.
The last point can be explained as follows: for $-1\leq \eta<0$ the
triplet $|1s2s;^{3}S\rangle$ state of the free space impurity
system, which corresponds to $\gamma\rightarrow\infty$,  is an
unbound state and has only one non-zero occupancy, that is
$\lambda_{00}^{-}={1\over2}$ \cite{en2}. In the case $s_{z}=\pm1$,
this  gives $S_{T}^{s_{z}=\pm1}=0$ (SR$=1$) and the corresponding
states $|\Psi_{ T}^{s_{z}=\pm1}\rangle$ must therefore be regarded
as  non-entangled. The situation is different  if one considers the
triplet state with $s_{z}=0$, namely: for
$\lambda_{00}^{-}={1\over2}$ its total wavefunction constitutes a
sum of two Slater determinants (SR$=2$) and this state has to be
regarded as an entangled state. Fig. \ref{fig:ooooodg} (b)
demonstrates, for the example of $\eta=-0.4$, how
 the occupancy $\lambda_{00}^{-}$ attains its asymptotic value ${1\over 2}$  with increasing
 $\gamma$. Except for the limits as   $\gamma\rightarrow0$ and  $\gamma\rightarrow\infty$,  where the dependence of the entanglement  on
 $\eta$ disappears, the triplet state with $s_{z}=\pm1$ is  generally  an entangled state. Nevertheless, as one can infer from  Fig.
\ref{fig:ooooodg} (a), the weakly entangled states with
$S_{T}^{s_{z}=\pm {1}}\approx0$ (SR$\approx 1$)
 are realized  for   finite values of $\gamma$ that  are  larger at bigger
   (negative)
   $\eta$.   For example, the vN entropy of the state $|\Psi_{ T}^{s_{z}=\pm1}\rangle$ of the system with $\eta=-0.4$
   starts to become vanishingly small already when $\gamma$ exceeds the value  $\gamma\approx
   20$.    Comparing the results of Fig. \ref{fig:odg} with
those of Fig. \ref{fig:ooooodg}, one can finally conclude that
for a given   $\eta$, the ranges of values of $\gamma$ in which the
entropies of the ground and  lowest triplet $S$ states make their
most rapid variations are nearly the same.

We close our discussion with Fig.  \ref{fig:eeeeeodg1}, which
compares the behaviour of the vN entropies of systems with a
donor impurity $\eta = 1$ and without impurity charge, for both
ground and triplet $S$ states.  The vN entropy  of the system with
$\eta = 1$
 exhibits  a monotonically increasing behaviour as $\gamma$
increases and goes to infinity as $\gamma\rightarrow\infty$,
similarly to the case for the impurity-free dots. The presence of
the donor impurity increases the entanglement for every $\gamma$
except for $\gamma=0$, which is more pronounced for the triplet
state than for the singlet one when $\gamma$ is large (the weak
confinement regime).
 It is well
known  that in the  case of an impurity-free dot, the angular
correlations carry the electrons at opposite sides of the centre of
the trap as $\gamma$ increases, i.e., the so-called linear Wigner
molecule is formed \cite{cios}. Being dependent  only on  the radial
coordinate, the donor impurity has thus an  effect mainly on the
radial correlations when the confinement becomes  weak. In this
regime, the increase in the entanglement resulting from the impurity
invasion comes thus mostly from an increase in radial correlations.
As can be inferred from Fig. \ref{fig:eeeeeodg1}, the larger is the
value of $\gamma$, the smaller is the change in the entanglement (in
a relative sense) produced by the coming of a donor impurity. The
impact of the donor impurity is therefore relatively small in the
weak confinement regime  in contrast to the acceptor impurity where
the opposite behaviour occurs.

\section{Conclusions}\label{summ}
 In conclusion, we carried out a comprehensive study of the entanglement properties of  two interacting electrons
 in the presence of Coulomb impurities in spherical harmonically shaped traps.
Our results showed  the dependencies of the vN entropy on  the dot
size and the  effective charge
 for both the ground state and the triplet $S$ states with the lowest energy.
From the results, it is apparent that the invasion of the charge
impurity dramatically influences the entanglement.
 The effect   is much more pronounced   when the impurity is negatively charged.
In such a case, when $\gamma$ is large enough,  the effect of
the confining potential is negligible  and  the entanglement entropy
approaches the value of the corresponding state of the free space
impurity
 system. As a general trend, we found that the closer
$\eta$   is to zero,  the larger is the value of $\gamma$
at which this occurs. On the other hand, when the confinement is
strong (small $\gamma$), the impact of the impurity is small  and
the entanglement approaches the value of the  impurity-free dots. It
turned out that the range of values of $\gamma$ around $\gamma=0$ in
which this occurs tends to increase when $\eta\rightarrow0^{-}$.
  In the
 limit as $\gamma\rightarrow\infty$, the vN entropy exhibits
 a discontinuity at the point $\eta=0$, since
  it tends to a constant value as
$\eta\rightarrow 0^{-}$, while at $\eta=0$, it tends to $\infty$.
 In
the case of a positively charged impurity, the vN entropy increases
monotonically as the dot size increases, similarly to the case for
the impurity-free dots. It was found  that the impact of the donor
impurity on the entanglement  is relatively small in the weak
confinement regime, i.e., where the Wigner molecule is formed.
 Except for the limit of infinitely
strong confinement ($\gamma\rightarrow0$), we found the lowest
singlet state to have generally higher entanglement than the lowest
triplet one with $s_{z}=\pm1$.

It would be desirable to gain a deeper insight into the properties
of the system  (\ref{hamff}) by  extending the current calculations
to  excited states. In particular,  as the confined systems of atoms
have important applications to modelling a variety of problems in
physics, it would be interesting to explore the entanglement of
the eigenstates of confined helium ions and check how the results
deviate from those for their free counterparts
\cite{4,10,11,12}.

\bibliography{aipsamp}

\begin{thebibliography}{99}
\bibitem{inf}Nielsen, N., and I. Chuang, Quantum Computation and Quantum
Information. Cambridge University Press, Cambridge, 2000.
\bibitem{inf1}Benenti, G., G. Casati, and G. Strini, Principles of
Quantum Computation and Information vols I and II. World Scientific, Singapore,
2007.
\bibitem{1} Amovilli, C., and N. March, Phys. Rev. A 69, 054302 (2004)
\bibitem{2} Ya\~{n}ez, R., A. Plastino, and J. Dehesa, Eur. Phys. J. D 56 (2010) 141
\bibitem{3} Majtey, A., A. Plastino,  and J. Dehesa,  J. Phys. A: Math. Theor. 45
(2012) 115309
\bibitem{3.1} Bouvrie, P. A.,  et al., Eur. Phys. J. D 66 (2012) 15

\bibitem{4} Manzano, D., et al.,  J. Phys. A Math. Theor. 43 (2010) 275301
\bibitem{5} Coe, J., A. Sudbery, and I. D'Amico, Phys. Rev. B 77
(2008) 205122
\bibitem{6} Ko\'{s}cik, P., and A. Okopi\'{n}ska,  Phys.
Lett. A 374 (2010) 3841
\bibitem{7} Ko\'{s}cik, P., Phys. Lett. A 375 (2011) 458
\bibitem{8} Ko\'{s}cik, P., and H. Hassanabadi, Few-Body Systems  52 (2012) 189-192
\bibitem{9} Nazmitdinov, R., et al., J. Phys. B: At. Mol. Opt. Phys. 45
(2012) 205503
\bibitem{10} Dehesa, J., et al., J. Phys. B: At. Mol. Opt.
Phys. 45 (2012) 015504
\bibitem{11} Benenti, G., S. Siccardi, and G. Strini, Eur. Phys. J.
D (2013)67, 83
\bibitem{12} Lin, Y., C. Lin, and Y. Ho, Phys. Rev. A 87 (2013)  022316
\bibitem{en1} Osenda, O., and P. Serra, Phys. Rev. A 75 (2007) 042331
\bibitem{en2} Osenda, O., and P. Serra, J. Phys. B At. Mol. Opt. Phys. 41 (2008) 065502
\bibitem{tishy}Tichy, M., F. Mintert, and A. Buchleitner 2011 J. Phys. B: At. Mol. Opt.
Phys. 44 192001


\bibitem{im} Genkin, M., and E. Lindroth, Phys. Rev. B 81 (2010)
125315

\bibitem{im2} Lee, C., C. C. Lam, and S. W. Gu, Phys. Rev. B 61 (2000) 10376


\bibitem{im5} Pandey, R., et al.,  J. Phys.: Condens.
Matter 16 (2004) 1769
\bibitem{im6} Kassim, H., J. Phys.: Condens. Matter 19 (2007) 036204

\bibitem{13} Banerjee, A., C. Kamal, A. Chowdhury, Phys.
Lett. A  350 (2006) 121–125


\bibitem{15} Laughlin, C., and S. I. Chu, J. Phys. A: Math. Theor. 42 (2009) 265004
\bibitem{16} Chakraborty S., and Y. K. Ho, Phys. Rev A 84 032515 (2011)
\bibitem{17} Aquino, N., A. Riveros, and J. Rivas-Silva, Phys. Lett. A 307 (2003) 326
\bibitem{18} Montgomery, H. Jr., N. Aquino, and A. Flores-Riveros, Phys.
Lett. A   374 (2010) 2044
\bibitem{19} Flores-Riveros, A., and A. Rodriguez-Contreras,  Phys. Lett. A 372 (2008) 6175

\bibitem{critical} Baker, D., E. Freund, R. N. Hill, and J. D. Morgan, Phys. Rev. A
41 (1990) 1241
\bibitem{hyleras} Hylleraas, E., Z. Phys. 54 (1929)  347
\bibitem{kaplan} Kaplan, W., Fourier--Legendre Series. §7.14 in Advanced Calculus,
4th ed., Addison-Wesley, Reading, MA, pp. 508--512, 1992.
\bibitem{vn1} Pa\v{s}kauskas, R., and L. You, Phys. Rev. A 64 (2001) 042310
\bibitem{ghirardi} Ghirardi, G., and L. Marinatto, Phys. Rev. A 70 (2004) 012109
\bibitem{rdm} Coleman, A., and V. Yukalov, Reduced Density Matrices,
Springer-Verlag, Berlin, 2000.
\bibitem{par} Grobe, R., K. Rz\c{a}\.{z}ewski, and J. Eberly, J. Phys. B 27 (1994) L503
\bibitem{vn31} Schliemann, J., et al. , Phys. Rev. A 64, 022303 (2001);
\bibitem{lin0} Buscemi, F., P. Bordone, and A. Bertoni, Phys. Rev. A 75 (2007) 032301
\bibitem{vn2} Plastino, A., D. Manzano, and J. S. Dehesa, Europhys. Lett. 86 (2009)
20005
\bibitem{vn3} Schr\"{o}ter, S., H. Friedrich, and
J. Madro\~{n}ero, Phys. Rev. A 87 (2013) 042507
\bibitem{hl} Hesse, M., and D. Baye, J. Phys. B: At. Mol. Opt. Phys. 32 (1999) 5605
\bibitem{cios} Cios{\l}owski, J., and M. Buchowiecki, J. Chem. Phys. 125 (2006) 064105.












\end{thebibliography}

\end{document}